\pdfoutput=1 
\documentclass{article}

\usepackage{arxiv}

\usepackage[utf8]{inputenc} 
\usepackage[T1]{fontenc}    
\usepackage{hyperref}
\hypersetup{colorlinks,allcolors=black}

\usepackage{url}            
\usepackage{booktabs}       
\usepackage{amsfonts}       
\usepackage{nicefrac}       
\usepackage{microtype}      
\usepackage{lipsum}
\usepackage{graphicx}
\graphicspath{ {./images/} }

\usepackage{graphicx}
\usepackage{physics}
\usepackage{amsmath}
\usepackage{breqn} 
\usepackage{xcolor} 
\usepackage{siunitx} 
\usepackage[british]{babel} 
\usepackage{tabularx} 
\usepackage{multirow}
\usepackage{algpseudocode}
\usepackage{algorithm}
\usepackage[utf8]{inputenc}
\DeclareUnicodeCharacter{2212}{-}
\usepackage{amsmath}

\usepackage{caption}
\usepackage{amsthm}
\usepackage{array}
\newcolumntype{P}[1]{>{\centering\arraybackslash}p{#1}}
\definecolor{maiblue}{rgb}{0, 0., 0.69}
\hypersetup{colorlinks=true, citecolor=maiblue, pdffitwindow=true, linkcolor=maiblue, urlcolor=maiblue}

\usepackage{hyperref}
\usepackage{color, colortbl}

\definecolor{Gray}{gray}{0.925}

\title{Quantum Splines for Non-Linear Approximations}

\author{
  Antonio Macaluso\\
  Department of Computer Science and Engineering \\
  University of Bologna, Italy\\
   \texttt{antonio.macaluso2@unibo.it} \\
   \And
    Luca Clissa \\
    Department of Physics and Astronomy\\
    University of Bologna, Italy\\
    Istituto Nazionale di Fisica Nucleare (INFN), Italy \\
    \texttt{luca.clissa2@unibo.it} \\
  \AND
      Stefano Lodi \\
    Department of Computer Science and Engineering \\
    University of Bologna, Italy\\
    \texttt{stefano.lodi@unibo.it} \\
    \And
    Claudio Sartori \\
    Department of Computer Science and Engineering \\
    University of Bologna, Italy\\
    \texttt{claudio.sartori@unibo.it} \\
}

\begin{document}
\maketitle

\begin{abstract}
Quantum Computing offers a new paradigm for efficient computing and many AI applications could benefit from its potential boost in performance.
However, the main limitation is the constraint to linear operations that hampers the representation of complex relationships in data.
In this work, we propose an efficient implementation of quantum splines for non-linear approximation. In particular, we first discuss possible parametrisations, and select the most convenient for exploiting the HHL algorithm to obtain the estimates of spline coefficients.
Then, we investigate QSpline performance as an evaluation routine for some of the most popular activation functions adopted in ML.
Finally, a detailed comparison with classical alternatives to the HHL is also presented.
\end{abstract}


\keywords{Quantum Computing \and Quantum AI \and Machine Learning \and Neural Networks \and HHL}


\maketitle

\section{Introduction}

Quantum Computing (QC) is an emerging field with the potential to revolutionise the world of information technology by leveraging quantum mechanics to endow quantum machines with tremendous computing power. This would enable the solution of problems not feasible to address with classical devices.
Of course, these premises are hugely appealing for many real-world applications, especially when coming to the adoption of Machine Learning and Deep Learning models in the domain of Artificial Intelligence.
However, the operations on quantum states are required to be linear and unitary under the laws of quantum mechanics \cite{nielsen2002quantum}.
This limitation deprives these techniques of their main strength, i.e., the ability to accurately describe complex non-linear relations.

\subsection{Background} \label{sec:background}

Spline functions are smoothing methods suitable for modelling the relationships between variables, typically adopted either as a visual aid in data exploration or for estimation purposes \cite{hastie01statisticallearning}.
The underpinning idea is to use linear models in which the input features are augmented with the so-called \textit{basis expansions}. These consist of transformations of the original variables and serve to introduce non-linearity.
Technically, splines are constructed by dividing the sample data into sub-intervals delimited by breakpoints, also referred to as knots.
A fixed degree polynomial is then fitted in each of the segments, thus resulting in a piecewise polynomial regression.
Formally, in the case of a $1$-dimensional input vector $\boldsymbol{x}$, we can express its relationship with a target variable $\boldsymbol{y}$ in terms of an order-$M$ spline with knots $\left\{\xi_k\right\}_{k=1, ..., K}$:
%
\begin{align}
\boldsymbol{y}_{n_{\text{obs}}\times1} 
 =  \boldsymbol{N}_{n_{\text{obs}}\times\left(M + K \right)} \boldsymbol{\theta}_{(M + K)\times1} + \boldsymbol{\epsilon}_{n_{\text{obs}}\times1} ,
\end{align}
where $\boldsymbol{\theta}$ is the vector of coefficients attached to the basis expansions, $n_{\text{obs}}$ is the sample size, $\boldsymbol{\epsilon}$ is a random error term and the design matrix $\boldsymbol{N}$ contains $M+K$ basis functions defined as follows:
\begin{align}
h_j(x) & = x^{j-1}, & j=1, \cdots, M 
\\
h_{M+k} & = \left( x - \xi_k\right)_+^{M-1}, & k=1, \cdots, K .
\end{align}
Notice that the formulation above includes $M$ basis functions that determine the order-$M$ polynomial to be fitted in each segment. The additional $K$ basis introduce continuity constraints on the spline and its derivatives up to order $M-2$.
The goal is then to find the optimal set of parameters $\boldsymbol{\theta}$ that minimises the residual sum of squares (\textit{RSS}), with a \textit{ridge regularisation} of the curvature acting as a roughness penalty:
\begin{align}\label{splinescore}
\text{Score} \left( \boldsymbol{\theta}, \eta \right) = \left( \boldsymbol{y} - \boldsymbol{N }\boldsymbol{\theta} \right)^T
\left( \boldsymbol{y} - \boldsymbol{N} \boldsymbol{\theta} \right) + \eta \boldsymbol{\theta}^T \boldsymbol{\Omega}_{(M+K) \times (M+K)} \boldsymbol{\theta}
,
\end{align} 
where $\boldsymbol{\Omega}$ is a diagonal matrix containing the partial second derivatives. The solution to \eqref{splinescore} can easily seen to be:
\begin{align} \label{eq:theta_hat}
\hat{\boldsymbol{\theta}} =  ( \boldsymbol{N}^T  \boldsymbol{N} + \eta \boldsymbol{\Omega} )^{-1} \boldsymbol{N}^T \boldsymbol{y} .
\end{align}

\subsection{Contribution}
In this work, we demonstrate how to adopt quantum splines for approximating several popular activation functions commonly employed in Neural Networks. 
In practice, we followed two different strategies. The $hybrid$ approach computes quantum estimates of the spline coefficients via the Harrow Hassidim Lloyd (HHL) quantum algorithm. Then a classical device is used to evaluate the activation functions. The $full$ quantum approach, instead, takes care of the evaluation process end-to-end, 
with an additional circuit that reads the HHL estimates and evaluates the function.
An in-depth discussion of the HHL algorithm efficiency with respect to classical alternatives is also treated.

\subsection{Related Works}

One of the major issues for building a quantum Neural Network is the implementation of a non-linear activation function on a quantum system. 
Indeed, the operations on quantum states are required to be linear and unitary under the laws of quantum mechanics, as discussed in \cite{nielsen2002quantum}.
The most promising attempt in this direction is described in \cite{cao2017quantum}, where the authors used the repeat-until-success technique to design a quantum circuit for reproducing a non-linear activation function. 
The biggest limitation is that this function requires input in the range $\left[0, \frac{\pi}{2}\right]$, which is a severe constraint for real-world problems. 
A step forward was made in \cite{hu2018towards}, where  domain restrictions were  removed and the activation gate parameters were learned during training.

Regarding matrix inversion on quantum devices, the HHL algorithm \cite{harrow2009quantum} is the standard reference. The limitation of using HHL in real applications was discussed in \cite{aaronson2015read}. Therefore, the best way to look at it is as a template for other quantum algorithms. 
In \cite{clader2013preconditioned} the authors study the advantage of HHL in electromagnetic scattering cross-section, showing that it can potentially achieve exponential speed-up to arbitrary specified problems. 

\section{Quantum Spline}
The final goal of this work is to investigate non-linear approximation by means of a quantum spline (QSpline). In particular, we want to demonstrate its applicability as an evaluation routine by fitting the spline to some widespread activation functions adopted in Neural Networks.
To this end, we consider \textit{relu, elu, tanh} and \textit{sigmoid} and we use a linear spline with no derivability constraints. Also, no roughness penalisation is applied since we would like to mimic the target function as closely as possible.
Regarding the choice of knots, 20 equally spaced breakpoints were selected over the interval $(-1, 1)$.

\subsection{Implementation} \label{sec:implementation}
While the formulation in terms of truncated basis functions described in Section \ref{sec:background} is conceptually simple, its numerical and computational properties are not very attractive.
For this reason, in practice we adopt a \textit{B-splines} parametrisation \cite{de1978practical}. This allows generating a block design matrix where the \textit{sparsity}
is constant and depends on the degree of the polynomial fitted in each local interval.
Given a sequence of knots $\xi_1, \xi_2, \cdots, \xi_K$, we fit a line in each interval
$\left[\xi_k, \xi_{k+1} \right]_{k = 1, \cdots, K-1}$
without derivability constraints. In matrix form:
\begin{equation} \label{eq:B-spline_lin_system}
\boldsymbol{\Tilde{y}} = \boldsymbol{S\beta} \rightarrow 
\begin{pmatrix}
\Tilde{y}_1 \\
\Tilde{y}_2 \\
\cdots \\ 
\Tilde{y}_{K}\\
\end{pmatrix}
=
\begin{pmatrix}
S_1 & 0 & \cdots & 0\\
0 & S_2 & \cdots & 0\\
\cdots & \cdots    & \cdots & \cdots\\ 
0 & 0 & \cdots & S_{K}\\
\end{pmatrix} 
\begin{pmatrix}
\beta_1 \\
\beta_2 \\
\cdots \\ 
\beta_{K}\\
\end{pmatrix} 
,
\end{equation}
where $\Tilde{y_k}$ contains the activation function evaluations in $\xi_k$ and $\xi_{k+1}$, $\beta_k$s are the spline coefficients and $\boldsymbol{S}_{(2K) \times (2K)}$ is a block diagonal matrix with each block $S_k$ that represents the basis expansions in the $k$-th interval.
Solving the linear system in Eq. \eqref{eq:B-spline_lin_system} requires using HHL to invert the matrix $\boldsymbol{S}$. 
Nonetheless, we can exploit the fact that the inverse of a block diagonal matrix is still block diagonal, with the correspondent inverse matrices in each block.
This implies we can solve $K$ $2 \times 2$ quantum linear systems 
$S_k \ket{\beta_k} = \ket{\Tilde{y_k}}$
instead of a single one for the entire function.
This little trick permits to overcome the practical limitations of the available quantum simulators, thus enabling the calculation of the spline coefficients 
through quantum simulations.

In particular, the computation of the full QSpline is performed in three steps. First, the HHL computes the spline coefficients for the $k$-th interval: 
\begin{equation}
S_k \ket{\beta_k} = \ket{\Tilde{y_k}} \xrightarrow{HHL} \ket{\beta_k} \simeq S_k^{-1} \ket{\Tilde{y_k}}.
\end{equation}
Second, $\ket{\beta_k}$ interacts with the quantum state encoding the input in the $k$-th interval via quantum interference. The scalar product between $\ket{\beta_k}$ and $\ket{x_{i,k}}$ is computed using the swap-test \cite{PhysRevLett.87.167902}:
\begin{align}
    \ket{\beta_k} \ket{x_{i,k}} \ket{0}  \xrightarrow{swap-test} \ket{e_1} \ket{e_2} \ket{f_{i,k}}.
\end{align}
At this point, the amplitudes of the quantum state $\ket{f_{i,k}}$ embed the estimate of the activation function evaluated in $x_{i,k}$. 
Third, $\ket{f_{i,k}}$ is measured to get the probability of state $\ket{0}$. This depends on the dot product between $\beta_k$ and $x_{i,k}$ as follows: 
\begin{align}
    \ket{f_{i,k}} = \sqrt{p_0}\ket{0} + \sqrt{p_1}\ket{1},
\end{align}
where: 
\begin{align} \label{eq:p_0}
    p_0 = \frac{1}{2}+\frac{|\langle \beta_k| x_{i,k} \rangle|^2}{2} = \frac{1}{2}+\frac{|f_{i,k}|^2}{2} .
\end{align}
Finally, the activation function estimate in correspondence of $x_{i,k}$ is retrieved by back-transforming Eq. \eqref{eq:p_0} to get $f_{i,k}$.

Notice that, the estimates are intrinsically bounded in the interval $[0,1]$ since they are encoded as the amplitude of a quantum state.
For this reason, the target activation functions are first scaled (i.e. $f_{i,k} \rightarrow f^*_{i,k} \subseteq [0,1]$). The QSpline is then trained on $f^*_{i,k}$, and the results are finally back-transformed to get the original $f_{i,k}$.

\section{Results}

\begin{figure}[ht]
    \centerline{\includegraphics[width=9cm]{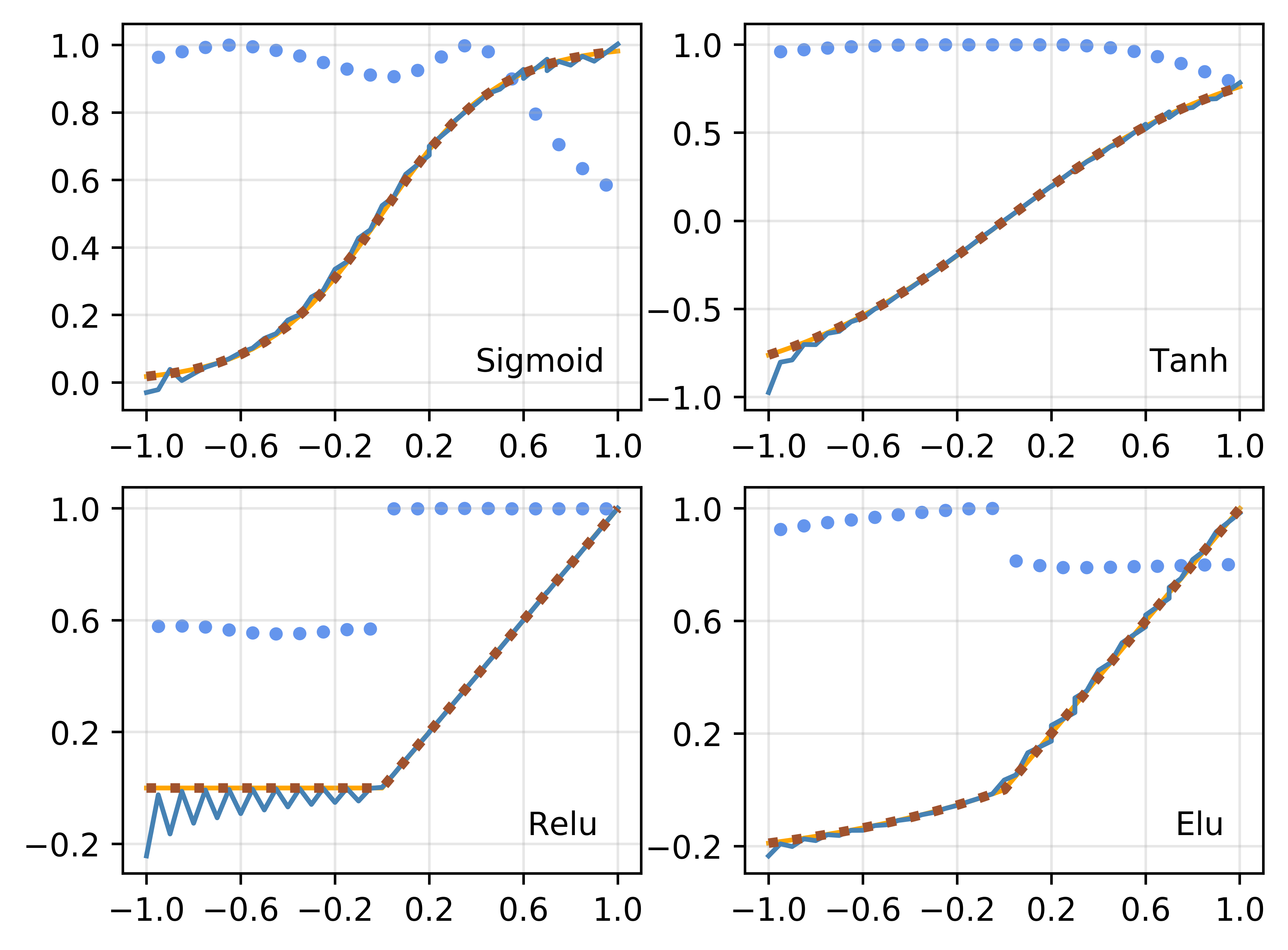}}
     \centerline{\includegraphics[width=9cm, height = .9cm]{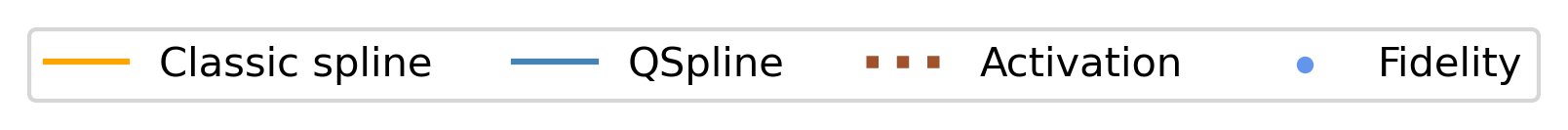}}
    \caption{Hybrid QSpline. }
    \label{fig:fit_results}
\end{figure}

The results of hybrid QSpline are illustrated in Fig. \ref{fig:fit_results}.

The quantum splines perform very well in reproducing the activation curves. The agreement is almost perfect for the sigmoid and the hyperbolic tangent, with slight deviations at the boundaries of the estimation interval. 
The situation is similar for Relu and Elu, although apparently some systematic error is introduced in the negative part of the $x$ domain.	
A possible explanation can be given by looking at the blue dots representing the fidelity of the quantum algorithm, that measures the discrepancy between the HHL output and the real solution of the system. In fact, larger deviations appear in areas where the fidelity is low, thus suggesting the HHL implementation may behave poorly in our setting.

\begin{figure}[ht]
     \centerline{
     \includegraphics[width=9cm]{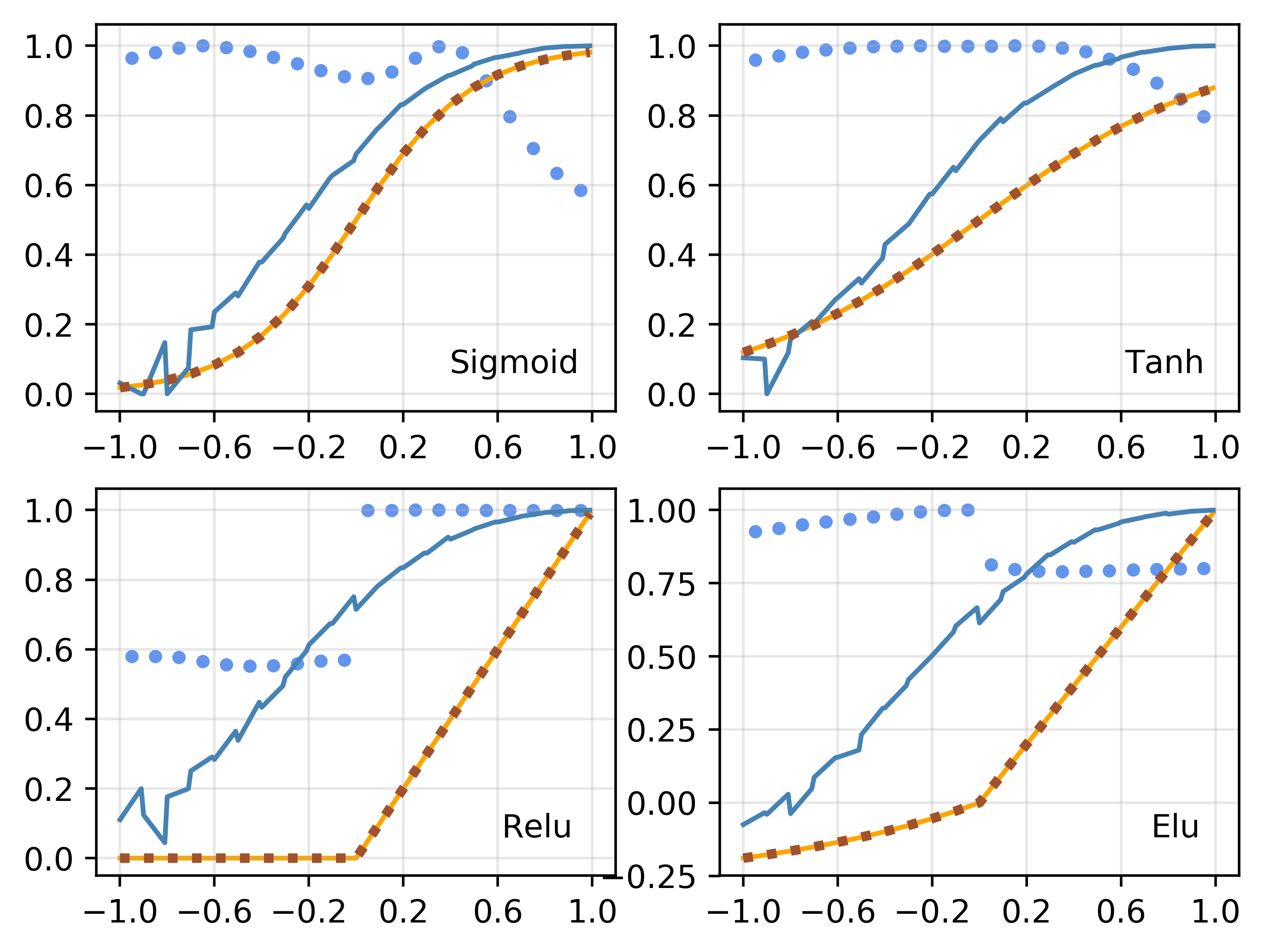}} 
    \centerline{\includegraphics[width=9cm, height = .9cm]{legend.png}}
    \caption{Full QSpline.}
    \label{fig:fit_quantum}
\end{figure}

A further attempt was then made with a subset of activation curves exploiting a \textit{full} quantum circuit.
The results are reported in Fig. \ref{fig:fit_quantum}.
In this case, the goodness of fit is definitely worse, with an almost systematic overestimate of the target curves. 
This may be due to the adoption of a second circuit for evaluation on top of the estimation one. Therefore, having two measurement phases introduce more uncertainty which is then propagated and produce a larger error.
Nonetheless, notice that the QSpline is likewise capable of reproducing the non-linear behaviour of the curves, which is anyway the most relevant characteristic under investigation.
For this reason, we interpret the results as promising, although they necessitate for more tuning.
A quantitative assessment of the goodness of fit of both strategies is reported in Table~\ref{tab:fit_results}.
\begin{table}[ht]
    \begin{center}
        \begin{tabular}[\linewidth]{c S[table-format=3.6] S[table-format=3.2] S[table-format=3.2] S[table-format=3.2]}
        \toprule
            \textbf{Activation} & \textbf{RSS}  & \textbf{RSS} & \textbf{RSS}  & \textbf{Average} \\
            \textbf{function} & \textbf{(classic)} & \textbf{(hybrid)} & \textbf{(quantum)} & \textbf{Fidelity} \\
            \toprule
            Sigmoid & \num{3.3e-05} & 0.01 & .75 & 0.90  \\
            Tanh & \num{1.2e-5} & 0.06 & 1.12 & 0.96\\
            Relu & {\hskip 1mm \num{7.6e-31}} & 0.14 & 8.16 & 0.78 \\
            Elu & \num{5.9e-07} & 0.12 & 7.06 & 0.88  \\[0.1cm]
            \bottomrule
        \end{tabular}
        \vspace{1em}
        {\caption{Approximation metrics. The table shows the Residual Sum of Squares (RSS) of both quantum and classical splines with respect to the true activation functions. Quantum approaches are indicated as \textit{hybrid} and \textit{quantum}. The average fidelity of the HHL is also reported.}
            \label{tab:fit_results}}
    \end{center}
\end{table}
\section{Computational Efficiency}
Here we discuss in details the computational cost of the quantum estimation phase, also drawing a theoretical comparison with the classical baselines.
In general, matrix inversion can be accomplished in polynomial time on classical devices \cite{burgisser2013algebraic, coppersmith1987matrix, strassen1969gaussian}. However, when several favourable assumptions hold it is possible to reduce computational costs. In particular, the \textit{Conjugate Gradient} algorithm \cite{ciliberto2018quantum, shewchuk1994introduction} allows solving a linear system with a complexity equal to $\mathcal{O}(s\sqrt{\kappa}n/\text{log}(\epsilon))$, where $n$ is the system size, $\kappa$ is the system condition number, $s$ the matrix sparsity (i.e. the maximum number of non-zero matrix elements of $\boldsymbol{A}$ in any given row or column), and $\epsilon$ is the desired error tolerance.

The reference quantum technique is the HHL algorithm. 
HHL is  a method for approximately preparing a quantum superposition of the form $\ket{\boldsymbol{x}}$, where $\boldsymbol{x}$ is the solution to a linear system $\mathbf{Ax = b}$, $\boldsymbol{A}$ is a hermitian design matrix and $\boldsymbol{b} $ is encoded in amplitudes of $\ket{\boldsymbol{b}}$. 
From a computational point of view, this requires an amount of time that grows roughly like $\mathcal{O}(s^2\kappa^2\text{log}(n)/\epsilon)$ (cfr. Table \ref{tab:HHL_efficiency} for a comparison between HHL and classical counterparts).
The algorithm scales logarithmically with respect to the size of the matrix, which means it has an exponential advantage when compared to classical alternatives. 
However, its complexity is polynomial in $s$ and $\kappa$, which means we have to introduce constraints on the condition number and the sparsity not to destroy the computational advantage of the HHL. This makes the previous comparison unfair since we cannot make assumptions about the design matrix in general.

\begin{table*}[ht]
\begin{center}
\begin{tabular}{c c c c c}
\toprule
\parbox{2cm}{\textbf{Gauss Jordan \\ elimination}}  & \textbf{Strassen} & \textbf{Coppersmith} & \parbox{2cm}{\textbf{Conjugate \\ Gradient}} & \textbf{HHL} \\[2mm]
\toprule
$\mathcal{O}(n^3)$ & $\mathcal{O}(n^{2.8})$ & $\mathcal{O}(n^{2.37})$ & $\mathcal{O}(sn\sqrt{\kappa}/\text{log}(\epsilon))$ & $\mathcal{O}(s^2\kappa^2\text{log}(n)/\epsilon)$ \\
\bottomrule
\end{tabular}
\vspace{1em}
{\caption{Comparison of algorithms computational costs.}
\label{tab:HHL_efficiency}}
\end{center}
\end{table*}
In our case, we can observe that an order-$M$ penalised spline has very desirable properties.
First of all, the ridge penalisation described in Eq. \eqref{eq:theta_hat} guarantees a lower condition number compared to the unconstrained design matrix. Specifically, Casella \cite{casella1985condition} showed that the higher the penalisation, the lower the condition number is. 
In fact, the condition number of a matrix, $\kappa(\boldsymbol{A})$, is equal to $\frac{|\lambda_{\textit{max}}(\boldsymbol{A}) + \eta|}{|\lambda_{\textit{min}}(\boldsymbol{A}) + \eta|} $, where $\lambda_{\textit{min}}$ and $\lambda_{\textit{max}}$ are the smallest and the biggest eigenvalues respectively, and $\eta$ is the amount of penalisation. 
Clearly, the more $\eta$ increases, the more it dominates the fraction, eventually tending to 1 for large penalisations.
Furthermore, regarding splines as piecewise polynomials defined on contiguous segments allows building a sparse design matrix $\boldsymbol{A}$, whose blocks describe the approximation in the corresponding intervals.
This implies that the sparsity amounts to the number of basis functions, $M+K$.
In light of the two properties above, splines are an ideal setting for an efficient application of the HHL algorithm.

\begin{figure}[ht]
    \centering
    \includegraphics[height = 6.3cm]{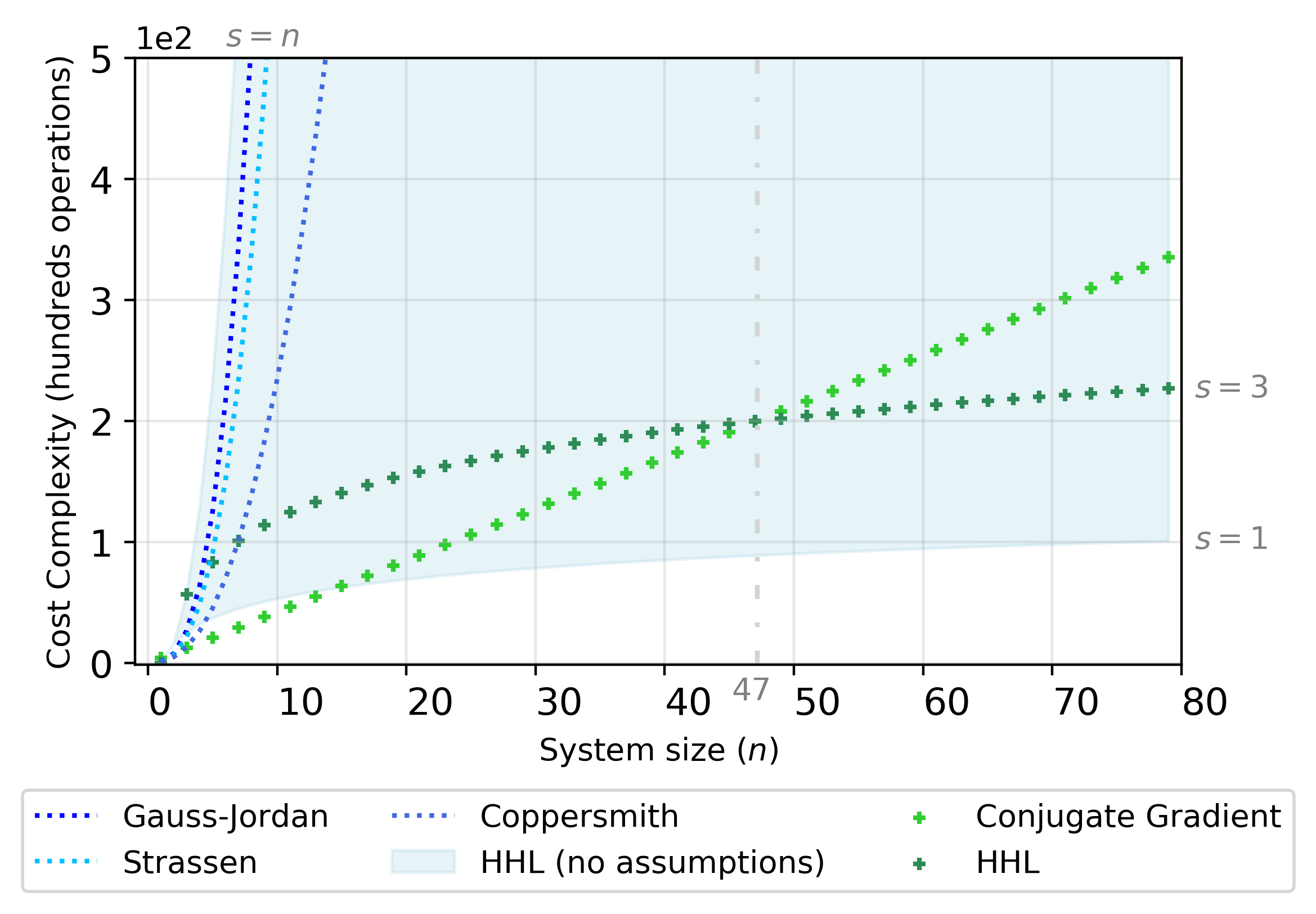}
    \caption{Cost complexity as a function of the size, $n$, of the $\boldsymbol{A}$ matrix. The green curves represent HHL and Conjugate Gradient for fixed $s, \kappa$, while the blue ones are referred to matrix inversion with no assumptions. The light blue shaded area illustrates the performance of HHL as the sparsity varies ($\kappa$ fixed).}
    \label{fig:HLL_efficiency}
\end{figure}

Figure \ref{fig:HLL_efficiency} illustrates a theoretical comparison of HHL computational costs with respect to the classical counterparts.
The green curves illustrate the computational costs (in hundreds of operations) of HHL and Conjugate Gradient for fixed sparsity and condition number. In particular, their values were chosen to reflect the case of natural splines with no intercept coefficient ($s=3$) and an approximately well-conditioned data matrix ($\kappa=2$). The HHL outperforms the Conjugate Gradient as the number of features becomes larger than $47$, that is quite frequent in \textit{Big Data} and \textit{Artificial Intelligence} applications, e.g. bioinformatics, natural language processing and computer vision.
A comparison of HHL with matrix inversion algorithms is also conducted when no assumptions are made. The light blue shaded are depicts the performance of HHL as $s$ varies ($\kappa$ fixed to $2$), while the curves using the blue palette describe classical alternatives. The advantage of HHL is evident as soon as some sparsity is introduced. However, we have also to take into account $\kappa$ and the higher costs of quantum computation to draw more solid conclusions.
Thus the efficiency boost due to the quantum technologies may foster the use of HHL for novel, classically unfeasible applications.
For instance, the computational burden of spline functions could be mitigated in this way, paving the way for future studies aimed at performing splines per se (e.g. multidimensional splines), and not just as a mere tool for evaluating non-linear functions. 

\section{Conclusion}\label{sec:conclusion}
In this work, we demonstrated the adoption of splines for approximating popular activation functions on a quantum device. The preliminary results were promising, although some tuning of the HHL implementation and the circuit for the full quantum approach is required. 
The quantum spline was able to reproduce the non-linearity of the curves, thus candidating this approach as a building block in the development of Quantum Neural Networks.

Also, we compared the computational complexity of the HHL algorithm against the most adopted classical counterparts, showing why splines may be an ideal setting for leveraging its advantages.

Future studies will be dedicated to improving the full quantum approach, with the possibilty of developing a novel and more stable implementation of the HHL routine.

\bibliographystyle{unsrt}  
\bibliography{references}

\end{document}